\documentclass[12pt]{iopart}

\usepackage{siunitx}
 \expandafter\let\csname equation*\endcsname\relax
  \expandafter\let\csname endequation*\endcsname\relax
\usepackage{amssymb}
\usepackage[fleqn]{amsmath}
\usepackage{iopams}  
\usepackage{graphicx}
\usepackage[breaklinks=true,colorlinks=true,linkcolor=blue,urlcolor=blue,citecolor=blue]{hyperref}
\usepackage{caption}
\usepackage{siunitx}

\usepackage{har2nat}   

\begin{document}

\title{Single--bubble dynamics in histotripsy and high--amplitude ultrasound: Modeling and validation}

\author[cor1]{Lauren Mancia$^{*,1,2}$, Mauro Rodriguez$^{3}$, Jonathan Sukovich$^{4}$, Zhen Xu$^{4}$ and Eric Johnsen$^{1}$}

\address{$^1$Department of Mechanical Engineering, University of Michigan, Ann Arbor, Michigan, USA}
\address{$^2$University of Michigan Medical School, Ann Arbor, Michigan, USA}
\address{$^3$Division of Engineering and Applied Science, California Institute of Technology, Pasadena, California, USA}
\address{$^4$Department of Biomedical Engineering, University of Michigan, Ann Arbor, MI, USA}

\ead{$^*$lamancha@umich.edu}

\begin{abstract}

A variety of approaches have been used to model the dynamics of a single, isolated bubble nucleated by a microsecond length high--amplitude ultrasound pulse (e.g.\, a histotripsy pulse). Until recently, the lack of single--bubble experimental radius vs.\ time data for bubble dynamics under a well--characterized driving pressure has limited model validation efforts. This study uses radius vs.\ time measurements of single, spherical histotripsy--nucleated bubbles in water [Wilson \textit{et al., Phys.\ Rev.\ E}, 2019, \textbf{99}, 043103] to quantitatively compare and validate a variety of bubble dynamics modeling approaches, including compressible and incompressible models as well as different thermal models. A strategy for inferring an analytic representation of histotripsy waveforms directly from experimental radius vs.\ time and cavitation threshold data is presented. We compare distributions of a calculated validation metric obtained for each model applied to $88$ experimental data sets.  There is minimal distinction ($< 1\%$) among the modeling approaches for compressibility and thermal effects considered in this study. These results suggest that our proposed strategy to infer the waveform, combined with simple models minimizing parametric uncertainty and computational resource demands accurately represent single--bubble dynamics in histotripsy, including at and near the maximum bubble radius. Remaining sources of parametric and model--based uncertainty are discussed.

\end{abstract}



\section{Introduction}

High--amplitude ultrasound therapies such as high intensity focused ultrasound (HIFU) and histotripsy are promising noninvasive treatments with a variety of clinical applications, including treatment of solid tumors, prostate pathologies, and biofilm infections \citep{khokhlova2015histotripsy,roberts2014development,bigelow2018histotripsy}.  Acoustically--driven inertial cavitation, the explosive growth and violent collapse of bubbles in response to acoustic forcing, is a known source of mechanical tissue damage during these treatments. In HIFU treatments, acoustic absorption causes tissue heating and subsequent thermal necrosis. Cavitation in the focal region can contribute to mechanical tissue damage in HIFU, but it is generally believed to be a secondary mechanism \citep{coussios2007role}. In contrast, histotripsy is a non--thermal focused ultrasound procedure that relies on targeted cavitation to homogenize soft tissue into acellular debris \citep{xu2004controlled,xu2005controlled,parsons2006pulsed,roberts2006pulsed}. Successful tissue fractionation in histotripsy requires the formation of a dense cloud of cavitation bubbles at the treatment focus \citep{parsons2007spatial,xu2005controlled}. The mechanisms for cloud formation and maintenance as well as the dynamics of cavitation bubbles within the cloud have been the subject of multiple experimental studies \citep{vlaisavljevich2014histotripsy,vlaisavljevich2015effects,vlaisavljevich2015effectsb,vlaisavljevich2016visualizing}. In intrinsic--threshold histotripsy, a single tensile phase of a waveform induces nuclei to grow at a threshold pressure that is intrinsic to the medium \citep{bader2019whom}. Within tens of microseconds, these nuclei believed to be nanometer--sized are observed to grow explosively into bubbles that are hundreds of microns in diameter \citep{maxwell2013probability,bader2019whom,mancia2020measurements}. Given these extreme conditions, experimental studies face limitations in temporal and spatial resolution as well as difficulties observing cavitation events in real tissues \citep{mancia2017predicting,vlaisavljevich2016visualizing}.

Numerical models for bubble dynamics under high--amplitude ultrasound forcing, including histotripsy, have been used to provide physical insight into phenomena occurring over time and length scales that are beyond current experimental resolution. Most modeling efforts to date attempt to simulate the dynamics of a single, spherical histotripsy bubble \citep{mancia2019modeling,mancia2017predicting,bader2018influence,bader2018influenceelastic,bader2016predicting}. These basic models offer the advantage of capturing fundamental bubble dynamics while using minimal computational resources. Demonstrating the adequacy of basic models could also simplify efforts to determine model parameters that are challenging to measure. For example, a study characterizing polyacrylamide gel under cavitation--relevant conditions found that a finite--deformation Kelvin--Voigt model was equivalent or superior to the more complex standard nonlinear solid model, which required consideration of non--equilibrium shear modulus as an additional unknown \citep{estrada2018high}. Moreover, identifying the essential physics needed to accurately model single bubbles remains an important step for advancing the development of higher--order methods \citep{rodriguez2019high} and bubble cloud models \citep{maeda2019bubble,ma2018numerical,fuster2011modelling} representative of histotripsy treatments. These methods require extensive computational resources and are not yet suitable for real--time treatment monitoring \citep{rossinelli201311}. Ultimately, treatment planning and monitoring will rely on efficient algorithms capable of predicting the onset of cavitation and on quantifying the extent of cavitation--induced tissue damage even in the simple case of a single, spherical bubble \citep{mancia2019modeling}. These models have demonstrated qualitative agreement with experiments \citep{vlaisavljevich2014histotripsy,vlaisavljevich2015effects,vlaisavljevich2015effectsb,vlaisavljevich2016visualizing}, but direct quantitative comparisons were limited by an inability to reliably isolate single bubbles under typical experimental conditions. 

Various models for bubble radial dynamics (compressible/incompressible), bubble contents (thermal effects), and driving pressure have been used in histotripsy simulations. Bubble dynamics models based on a Rayleigh--Plesset formulation have been validated by experiments in sonoluminescence \citep{ketterling1998experimental} and laser cavitation \citep{lam2016dynamical}. However, quantitative agreement of these models with high--amplitude ultrasound experiments has yet to be achieved, in part due to the lack of relevant experimental data to date. For radial dynamics, the Keller--Miksis \citep{keller1980bubble} and Gilmore--Akulichev equations \citep{gilmore1952growth,akulichev1967pulsations,church1989theoretical} are extensions of the Rayleigh--Plesset \citep{plesset1949dynamics} equation accounting for weak compressibility effects. Some authors recommend using the Gilmore model for cases of large amplitude bubble oscillations \citep{vokurka1986comparison}, while others argue that enthalpy formulations of the Keller--Miksis equation are more accurate \citep{prosperetti1986bubble}. Treatments of thermal effects also vary considerably, with most studies adopting a polytropic approximation for gas pressure inside the bubble \citep{vlaisavljevich2014histotripsy,vlaisavljevich2015effects,vlaisavljevich2015effectsb,vlaisavljevich2016visualizing,bader2016predicting,mancia2017predicting}. A more complete treatment of thermal effects \citep{prosperetti1988nonlinear} was used previously to study the effect of water temperature on the histotripsy intrinsic threshold \citep{vlaisavljevich2016effects}. More recently, a study of tissue--selective effects in histotripsy \citep{mancia2019modeling} used a simplified version of this model considering temperature variation inside the bubble while assuming surroundings remain at a constant temperature (i.e. the cold--medium assumption) \citep{prosperetti1991thermal,estrada2018high}. However, histotripsy is considered a nonthermal therapy \citep{roberts2006pulsed}, so it is also reasonable to assume isothermal conditions \citep{mancia2017predicting}. Model forcing waveforms have included a Gaussian pulse envelope \citep{maxwell2013probability} and half--cycle tensile pulses \citep{vlaisavljevich2014histotripsy,vlaisavljevich2015effects,vlaisavljevich2015effectsb,vlaisavljevich2016visualizing,mancia2017predicting,mancia2019modeling} which are scaled by the peak negative pressure. A complete histotripsy waveform was also used in bubble dynamics simulations \citep{bader2016predicting}, though this approach is not necessarily superior due to uncertainty in the acoustic field at the instant and location of cavitation inception.

Comparisons between histotripsy experiments and modeling results have featured in studies of the intrinsic cavitation threshold \citep{vlaisavljevich2014histotripsy,vlaisavljevich2015effects}, bubble behavior \citep{vlaisavljevich2015effectsb,vlaisavljevich2016visualizing}, and cavitation--induced tissue damage \citep{mancia2017predicting,mancia2019modeling}.  To date, these comparisons have been general and largely qualitative.  Even a recent study comparing maximum radii obtained in histotripsy bubble behavior experiments with model--predicted maximum radii \citep{bader2016predicting} is limited by its use of experimental radius vs.\ time data which includes cloud cavitation events in which bubbles interact with each other, thus modifying the local pressure \citep{vlaisavljevich2015effectsb}. Robust quantitative validation of single--bubble models requires experimental observation and measurement of a single histotripsy--nucleated bubble driven by a known pressure forcing from inception through collapse. Only recently have advances in experimental methods \citep{wilson2019comparative,sukovich2020cost} offered the possibility of validating existing single--bubble models for acoustic cavitation dynamics under histotripsy--type forcing. Validation of models for single--bubble 
dynamics in histotripsy has been limited by a lack of information about the driving pressure due to the presence of many bubbles distorting the acoustic field. The development of a single--cycle waveform \citep{wilson2019comparative} and use of specialized high--speed videography permits, for the first time, a validation of histotripsy single--bubble modeling assumptions. 

Based on experimental measurements (waveform pressure and $R(t)$ images), this study presents a numerical strategy for accurately modeling single--bubble dynamics in histotripsy and related high--amplitude ultrasound therapies.  For this validation study, we consider modeling approaches and experiments in water to avoid introduction of additional model--based and parametric uncertainties needed to account for the viscoelastic response of tissue. We quantitatively compare different histotripsy modeling approaches to experimental radius vs.\ time data obtained from single--bubble nucleation events in water \citep{wilson2019comparative}. Section \ref{sec:methods} presents single--bubble experiments in brief, and then provides an overview of models for compressibility and thermal effects used to simulate histotripsy bubble dynamics.  Subsequently, Section \ref{sec:results} validates a common analytic approximation of the histotripsy waveform and applies the methods and validation metric presented in a study of cavitation nuclei \citep{mancia2020measurements} to evaluate select models for compressibility and thermal effects assuming that the primary parametric uncertainty for acoustic cavitation in water lies in the initial radius or intrinsic nucleus size. Representative radius vs.\ time experimental data sets are used for comparison to simulation results obtained with different models. Finally, Section \ref{sec:discuss} presents a quantitative summary of initial radius statistics and validation metrics for each treatment of compressibility and thermal effects. Remaining parametric and model--based uncertainties in simulating single--bubble dynamics in histotripsy are described.

\section{Methods}
\label{sec:methods}

\subsection{Single--bubble Experiments}

The present experimental data was originally presented in a study comparing laser and acoustic cavitation in gels and water; further details regarding experimental methods can be found in this original work \citep{wilson2019comparative}. Water was deionized, filtered to 2 $\mu$m, and degassed to 4 kPa. Experiments were performed in a spherical acoustic array containing 16 focused transducer elements with a central frequency of 1 MHz. Single bubbles were nucleated using a 1.5--cycle histotripsy pulse with a single rarefactional pressure half--cycle. Images of the bubbles through a single cycle of growth and collapse were obtained using a high--speed camera with a multi--flash--per--camera--exposure technique \citep{sukovich2020cost}. To minimize the likelihood of nucleating multiple bubbles at the focal site, an empirical threshold study was performed prior to experiments. The acoustic rarefactional pressure was varied such that the probability of generating cavitation at the focus was $50\%$, as in previous studies of the histotripsy intrinsic threshold \citep{vlaisavljevich2014histotripsy,vlaisavljevich2015effects,vlaisavljevich2016effects}. This procedure resulted in a peak pressure of $-24$ MPa. The dimensional and scaled radius vs.\ time data for 88 acoustically--nucleated single--bubble experiments in water are shown in Mancia et al.\ (2020). Each data set is comprised of $15$ -- $25$ data points. The magnitudes of spatial and temporal resolution uncertainties in the experiments were $4.3$ microns and $\leq$1.25 microseconds, respectively.

\subsection{Theoretical Model}

In histotripsy and related high--amplitude ultrasound therapies, bubbles are assumed to arise from preexisting or intrinsic nuclei \citep{mancia2020measurements}. The high tension of the present waveform causes explosive growth and violent collapse of a single spherical bubble, in which case compressible, and possibly thermal, effects are expected to affect the dynamics. In this study, we consider varying fidelities in representing compressibility and thermal effects in our modeling of the dynamics of a single bubble in water subjected to a histotripsy pulse; these results can readily be extended to other high-amplitude ultrasound waveforms. While viscous and surface tension effects are also expected to play a role in the process, uncertainties pertaining to these effects are negligible in the present context because those properties of water are well characterized. The following subsections present a series of three models for compressibility effects and four models for thermal effects. Discussion of the derivation and wider applicability of each modeling approach can be found in the cited literature.
 
\subsubsection{Compressibility Effects}
 Assuming spherical symmetry, a homobaric gas, and an incompressible near field, the equations of motion can be integrated to yield an ordinary differential equation describing the time evolution of the bubble radius in water. If the bubble wall velocity is sufficiently large compared to the liquid sound speed, liquid compressibility gives rise to the radiation of acoustic waves. We represent these compressibility effects using three models of varying fidelity, written in  the following general form:
\begin{equation}\label{eq:RadEq}
\begin{split}
\left(1 -\frac{\dot{R}}{c_{\infty}}\right)R&\ddot{R}+\frac{3}{2}
 \left(1 - \frac{\dot{R}}{3c_{\infty}}\right)\dot{R}^2 
 =\frac{1}{\rho_{\infty}}\left(1 +\frac{\dot{R}}{c_{\infty}}+
 \frac{R}{c_{\infty}}\frac{d}{dt}\right)(\Psi(t,R)-p_{ac}(t)), \\
 & \Psi(t,R)= 
\begin{cases}
p-p_{\infty},\text{ } c_{\infty} \rightarrow \infty, & \text{Rayleigh--Plesset}     
    \\p-p_{\infty}, & \text{Keller--Miksis with Pressure}
     \\ h_B, &\text{Keller--Miksis with Enthalpy}, 
\end{cases}
 \end{split}
\end{equation}

\noindent
where $\Psi(t,R)$ corresponds to three different treatments of compressibility: Rayleigh--Plesset, Keller--Miksis with pressure, and Keller--Miksis with enthalpy. All three models depend on pressure evaluated at the bubble wall: $p = p_B -\frac{2S}{R} - 4\mu\frac{\dot{R}}{R}$, where the internal bubble pressure, $p_B$ is defined in Sect.\ \ref{sec:thermal} on thermal effects, $S$ is surface tension, and $\mu$ is dynamic viscosity. The Keller--Miksis with enthalpy formulation also depends on the pressure through the enthalpy evaluated at the bubble wall, $h_B$:
\begin{equation}
\begin{aligned}\label{eq:Enthalpy}
h_B = \frac{n}{n-1}\left(p_{\infty} + B\right)\Biggl[\left(\frac{p + B}{p_{\infty} + B}\right)^{\frac{n-1}{n}} - 1\Biggr].
\end{aligned}
\end{equation}
\noindent
Constant parameters $n$ and $B$ for the modified Tait equation of state are defined for water as in Prosperetti and Lezzi (1986). These constants as well as the density, $\rho_{\infty}$, static fluid pressure, $p_{\infty}$, constant sound speed, $c_{\infty}$, surface tension, $S$, and viscosity, $\mu$, of surrounding water are given in Table \ref{table:Constants}. The acoustic forcing pressure,  $p_{ac}(t)$, experienced in the focal region is described by Eq.\ \ref{eq:Pac}, and its determination is discussed in Sect. \ref{sec:waveform}. The Rayleigh--Plesset equation \citep{rayleigh1917viii,plesset1949dynamics} assumes fully incompressible surroundings while both the Keller--Miksis with pressure and Keller--Miksis with enthalpy equations account for first--order compressibility effects. In the limit of small bubble wall velocity relative to the liquid sound speed, all models converge to the incompressible solution; discrepancies between models become manifest as the bubble reaches wall velocities that are no longer negligible compared to the liquid sound speed. The pressure--based Keller--Miksis model was previously shown to be slightly inferior to its enthalpy formulation \citep{prosperetti1986bubble}. The same study suggested that the Gilmore (1952) and Herring (1941) models are also inferior to the Keller--Miksis with enthalpy formulation, so these models are not considered here. A distinguishing feature of the Gilmore equation is its inclusion of variable sound speed, which in theory makes it applicable up to larger Mach numbers ($\dot{R}/c \lesssim 2.2$) than the constant sound speed Keller--Miksis equations ($\dot{R}/c \lesssim 1$) \citep{zilonova2018bubble}. In practice, however, the Mach numbers reached by the histotripsy bubble wall remain well within the range of validity for Keller--Miksis models until the instant of bubble collapse to a nanoscale minimum radius.  Limitations in simulating the complex physics at histotripsy bubble collapse are inherent to all available lower--order models and are discussed in Sect.\ \ref{sec:discuss}.

\subsubsection{Thermal Effects}
\label{sec:thermal}

The thermodynamics couple to the  dynamics via the bubble pressure, given by: 
\begin{align} \label{eq:pB}
\dot{p}_B=\frac{3}{R}\left((\kappa -1)K\frac{\partial T}{\partial r}\bigg|_{R}-\kappa p_B\dot{R}\right),
\end{align}
where $T(r,t)$ is the temperature field, $K$ is the gas thermal conductivity equal to $K_AT + K_B$ where $K_A$ and $K_B$ are empirically determined constants \citep{prosperetti1988nonlinear}, and $\kappa$ is the specific heats ratio of the gas. We consider four thermal approaches, whose representation of the heat flux at the bubble wall dictate model fidelity: full thermal model, cold--liquid assumption, and polytropic approximation (isothermal and adiabatic). In the full thermal model \citep{prosperetti1988nonlinear,kamath1993theoretical,barajas2017effects,
warnez2015numerical}, the energy equation is solved inside the bubble:
\begin{equation} \label{eq:kappa}
\frac{\kappa}{\kappa-1}\frac{p_B}{T}\left[\frac{\partial T}{\partial t}+\frac{1}{\kappa p_B}\left((\kappa -1)K\frac{\partial T}{\partial r}-\frac{r\dot{p}_B}{3}\right)\frac{\partial T}{\partial r}\right]
=\dot{p}_B + \frac{1}{r^2}\frac{\partial}{\partial r}\left(r^2K\frac{\partial T}{\partial r}\right),
\end{equation}

\noindent
and outside the bubble:
\begin{equation} \label{eq:Tm}  
\frac{\partial T_M}{\partial t}+\frac{R^{2}\dot{R}}{r^{2}}\frac{\partial T_M}{\partial r}=D_M \nabla^2T_M+\frac{12\mu}{\rho_{\infty}C_p}\left(\frac{R^2\dot{R}}{r^3}\right)^2,
\end{equation}
where $T_M(t,r)$ is the temperature, $C_p$, is the specific heat, $D_M = K_M/(\rho_{\infty}C_p)$ is the thermal diffusivity, and $K_M$ is the thermal conductivity of the surrounding liquid.  These constants are given for water in Table \ref{table:Constants}.  Boundary conditions are prescribed for the center of the bubble and far from the bubble: $\nabla T=0$ at $r=0$ and $T_{M}\rightarrow T_{\infty}$ as $r\rightarrow L$, where $L \gg R$ is the arbitrary boundary of the domain. Boundary conditions at the bubble--liquid interface couple the internal bubble temperature to the temperature field in the surroundings: $T_M(r,t)$: $T|_{r=R}=T_M|_{r=R}$ and $K_{r=R}\frac{\partial T}{\partial r}|_{r=R} = K_{M}\frac{\partial T_M}{\partial r}|_{r=R}$. The cold--liquid assumption assumes that the surrounding medium maintains a constant temperature, such that the energy equation in the liquid no longer needs to be solved and $T_M(R) = T_{\infty}$ \citep{prosperetti1991thermal}. The polytropic model \citep{noltingk1950cavitation,akulichev1967pulsations,keller1980bubble,apfel1981acoustic} pertains to either (i) an isothermal process where $\kappa = 1$ or (ii) an adiabatic process where the heat flux at the wall is zero and $\kappa$ is the specific heats ratio; in both cases, the first term on the right--hand--side of Eq.\ \ref{eq:pB} is zero. For a polytropic gas, Eq.\ \ref{eq:pB} is equivalent to:

\begin{align}\label{eq:poly}
p_B = p_0\left(\frac{R_0}{R}\right)^{3\kappa}, \text{ } \kappa = 
\begin{cases}
1, & \text{Isothermal}\\     
1.4, & \text{Adiabatic},
\end{cases}
\end{align}

\noindent
where $p_0 = p_{\infty} + 2S/R_0 + p_v$ is the initial pressure inside the bubble, and $p_v$ is the vapor pressure of water at $25$ $^{\circ}$C. Both conditions have been assumed in histotripsy simulations \citep{vlaisavljevich2015effects,vlaisavljevich2015effectsb,vlaisavljevich2016visualizing}, but the isothermal assumption was shown to more closely approximate the results of a full heat transfer model \citep{mancia2017predicting}.

\subsection{Problem Setup}

\begin{table}
\centering
\captionsetup{justification=centering}
\caption{Constant Parameters}
\begin{tabular}{c c c c} 
 \hline\hline
 Parameter & Value & Parameter & Value\\
\hline 
  $p_{\infty}$ & \SI{101.325} {\kPa} &  $D_M$ & \SI{1.41e-7}{\meter \squared \per {\second}}\\
  $c_{\infty}$ & \SI{1497} {\meter \per {\second}}& $B$ & \SI{304.91}{\MPa}\\
  $T_{\infty}$ & \SI{25} {\celsius} &  $n$ & 7.15\\ 
 $p_V$ & \SI{3.169}{\kPa} &  $p_A$ & \SI{-24} {\MPa}\\ 
  $S$ & \SI{0.072} {\N \meter} &  $\eta$ & 3.7\\  
  $\mu$ & \SI{0.001} {\Pa \second} &  $f$ & \SI{1} {\MHz}\\
  $\kappa$ & 1.4 &  $\delta$ & \SI{5} {\micro\second}\\
  $C_p$ & \SI{4181}{\J \per {\kilogram \K}} \\
  $K_A$ & \SI{5.28e-5}{\W \per {\meter \K \squared}}\\
  $K_B$ &  \SI{1.165e-2}{\W \per {\meter \K}}\\
  $K_M$ & \SI{0.55}{\W \per {\meter \K}}\\ 
  \hline\hline 
\end{tabular}
\label{table:Constants}
\end{table}

The equations are nondimensionalized using the initial bubble radius, $R_0$, water density, $\rho_{\infty}$, and far--field temperature $T_{\infty}$ \citep{barajas2017effects}. The static fluid pressure, $p_{\infty}$, is used to define a characteristic speed, $u_c = \sqrt{p_{\infty}/{\rho_{\infty}}}$, and dimensionless 
parameters: Reynolds number, Re $= \rho_{\infty} u_cR_0/\mu$, Weber number, We $=p_{\infty}R_0/2S
$, dimensionless sound speed, $C=c_{\infty}/u_c$, Fourier number, $F_0 = D_M/u_cR_0$, Brinkman number, $Br = u_c^2/C_pT_{\infty}$, and $\chi = 
T_{\infty}K_M/p_{\infty}R_0u_c$. A variable--step, variable--order solver (MATLAB \textit{ode15s}) is used for time marching \citep{shampine1997matlab,shampine1999solving}. Equations are 
integrated over a dimensional time span of $t = [0,50]$ in 
microseconds; results are time--shifted so that the maximum bubble radius occurs at $t = 0$. Using numerical methods described by Warnez and Johnsen (2015), the spatial derivatives in the energy equation are discretized on a mesh of $N_s
+1$ points in $r$-space \citep{prosperetti1988nonlinear} inside and outside of the 
bubble and computed using a spectral collocation method \citep{warnez2015numerical}. Results are sufficiently converged when simulations use $N_s=30$ 
points inside and outside of the bubble. A more detailed treatment of the derivation and numerical implementation of this model can be found in the literature \citep{prosperetti1988nonlinear,kamath1993theoretical,barajas2017effects,
warnez2015numerical}.

\section{Results}
\label{sec:results}

We first present the validation metric used to quantify agreement between modeling and experimental results. Next, experimental bubble radius vs.\ time measurements are used to infer the acoustic forcing experienced by a single bubble in the focal region.  Subsequently, numerical simulations are performed combining modeling approaches for compressibility (Eq.\ \ref{eq:RadEq} and \ref{eq:Enthalpy}) and thermal (Eqs.\ \ref{eq:pB} -- \ref{eq:poly}) effects.  To review, for compressibility effects, we consider the Rayleigh--Plesset equation, Keller--Miksis equation with pressure, and Keller--Miksis equation with enthalpy. Thermal effects are modeled using adiabatic and isothermal polytropic approximations, the cold--liquid assumption, and a full thermal model. For quantitative comparison of these different modeling approaches, we present the optimized validation metric distributions and $R_0$ distributions obtained by applying select models to each of the $88$ experimental data sets. Summary statistics for all combined compressibility and thermal models are given in Sect.\ \ref{sec:discuss}.

\subsection{Validation Metric}
\label{sec:metric}

\begin{figure}
  \includegraphics[width=0.5\linewidth]{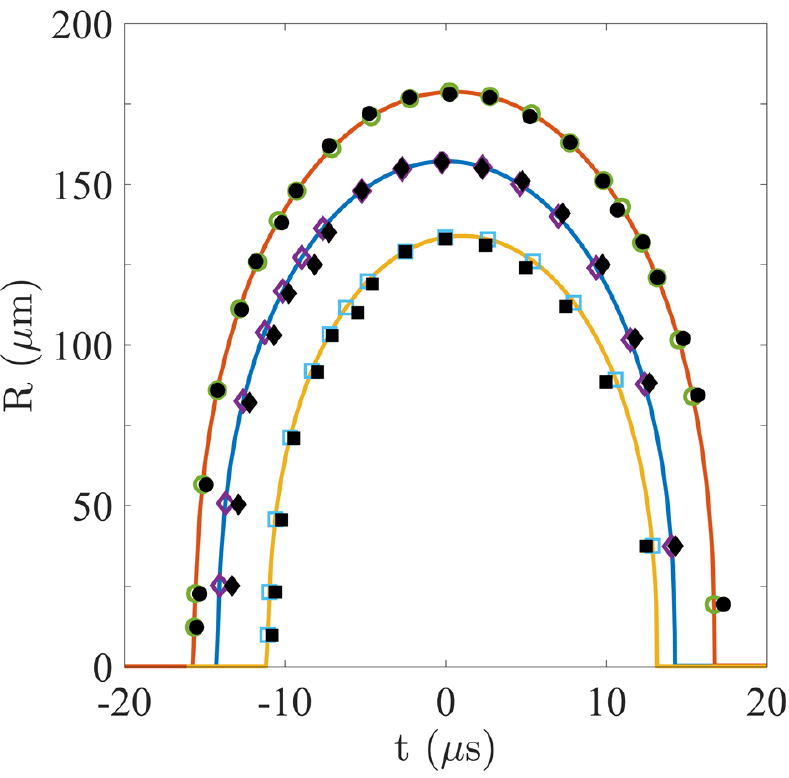}
  \centering
  \caption{Three representative experimental data sets are shown as black solid markers: $R_{max}^{\mu - \sigma}$ (squares), $R_{max}^{\mu}$ (diamonds), and $R_{max}^{\mu + \sigma}$ (circles). The traces through each data set are optimal simulations with nearest neighbor points identified by open markers.}
  \label{fig:VM}
\end{figure}

The validation procedure used to compare models for compressibility and thermal effects follows that of Mancia et al.\ (2020). This previous study identified the  radius used to initialize simulations (nucleus size, $R_0$) as the greatest source of parametric uncertainty in single--bubble experiments. For each experimental data set, numerical simulations are performed for a series of $R_0$ values with the intent of finding the $R_0$ that produces the best fit of simulation to data. The lower bound $R_0$ for all simulations was shown to be $2.32$ nm using the Blake threshold \citep{mancia2020measurements}. The simulation points closest to experimental points in each data set are then identified using a nearest neighbors algorithm with a standardized Euclidean distance metric. A nucleus size distribution is constructed for each modeling approach using $R_0$ values that optimize the normalized root--mean--squared (rms) error (NRMSE) between the simulation nearest neighbors and corresponding experimental points in each data set. The NRMSE validation metric ranges from $-\infty$ (worst fit) to $1.00$ (optimal fit), with most values falling between $0.88$ and $0.99$ in the present study. This procedure is illustrated in Figure \ref{fig:VM}, where three representative experimental data sets are shown as black solid markers. The traces through each data set are optimized simulations obtained using the Keller--Miksis equation with pressure and the cold liquid assumption. Nearest neighbor points on each simulation trace are identified with open markers. These representative radius vs.\ time data sets are selected because they have maximum radii near the mean maximum radius, $R_{max}^{\mu}$, of all data sets and approximately one standard deviation larger and smaller than this value, $R_{max}^{\mu \pm \sigma}$. In Figure \ref{fig:VM}, the $R_{max}^{\mu - \sigma}$ (squares), $R_{max}^{\mu}$ (diamonds), and $R_{max}^{\mu + \sigma}$ (circles) data sets, the optimal NRMSEs (with corresponding optimal $R_0$ for this model) are $0.969$ ($2.42$ nm), $0.979$ ($2.78$ nm), and $0.984$ ($3.52$ nm), respectively.

\subsection{Acoustic Waveform}
\label{sec:waveform}

Histotripsy transducers produce pressure waveforms with high--amplitude tensile and compressive components. A single acoustic cycle histotripsy ultrasound pulse with a high--amplitude tensile phase is typically used to generate cavitation through the intrinsic threshold approach \citep{maxwell2013probability}. Precise measurement of the forcing pressure as a function of time is challenging because these high--amplitude pressure waves ultimately produce destructive cavitation at the hydrophone tip \citep{vlaisavljevich2015effects,bader2016predicting}.  Additionally, existing bubbles and intervening material can distort the original waveform. Given these challenges, various representations of a histotripsy forcing waveform have been used in simulations. For example, a study of the cavitation threshold in tissues used a Gaussian pulse envelope \citep{maxwell2013probability}. An averaged waveform constructed from measured shock scattering histotripsy pulses was also used in bubble dynamics simulations \citep{bader2016predicting}.  In addition, multiple previous studies adopted a half--cycle tensile pulse fitted to the peak negative pressure portion of a raw histotripsy waveform \citep{vlaisavljevich2014histotripsy,vlaisavljevich2015effects,vlaisavljevich2015effectsb,vlaisavljevich2016visualizing,mancia2017predicting,mancia2019modeling,bader2018influence}.  
It is not obvious on first inspection that this tensile half--cycle is representative of the acoustic forcing experienced by a single cavitation bubble in the focal region; however, a strong case for the validity of this approximation can be made using the single--bubble data now available and the concept of an intrinsic cavitation threshold.

\begin{figure}[h]
\centering
\includegraphics[width=6.0 in]{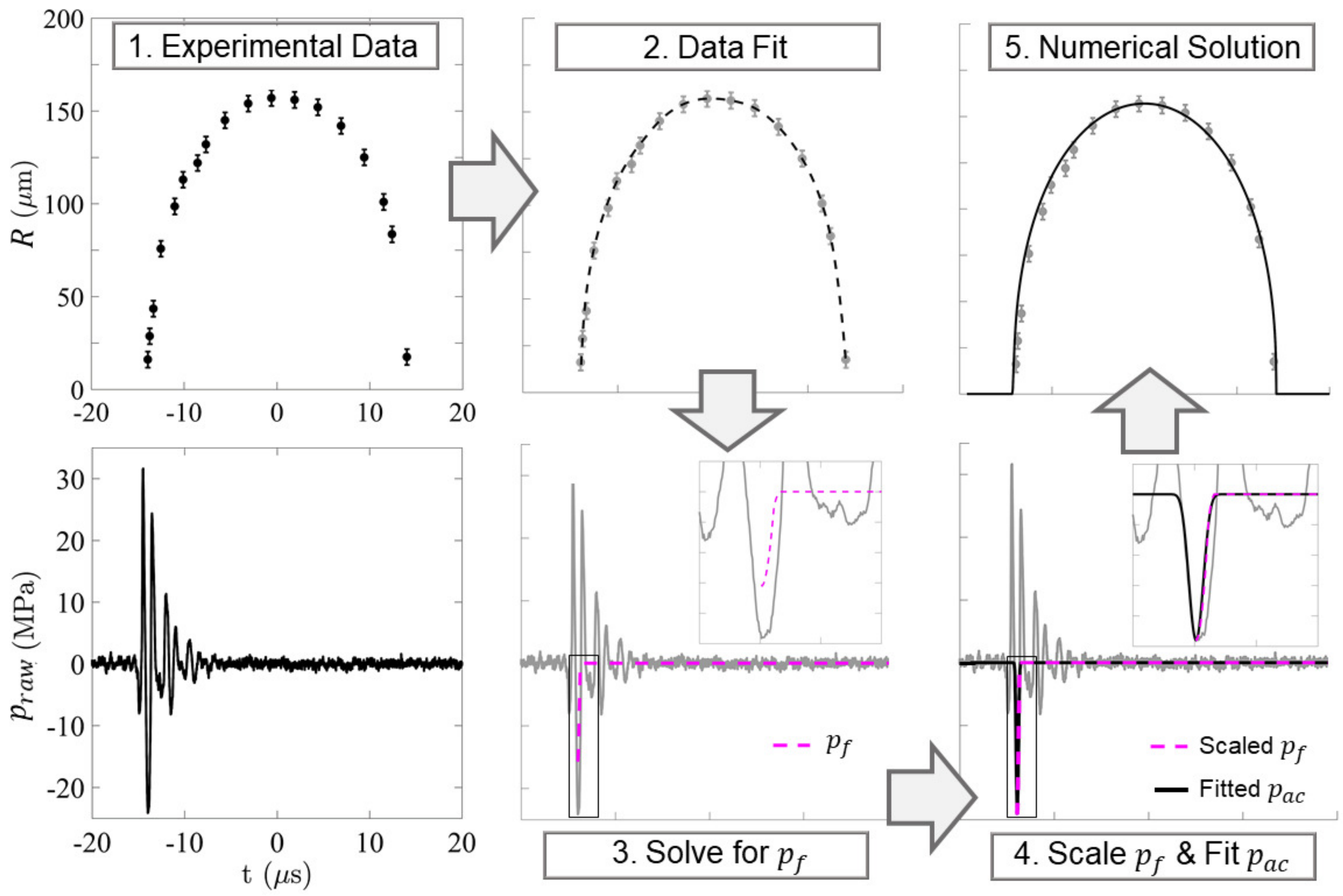}
  \caption{Inference of analytic driving pressure waveform: (Panel 1) Representative experimental data set with spatial resolution error bars (top) and raw histotripsy waveform with a peak pressure of $-24$ MPa (bottom). (Panel 2) Smooth curve fit through bubble radius vs.\ time data set using cubic splines method. (Panel 3) Assume curve fit solves the Rayleigh--Plesset equation and solve Eq.\ \ref{eq:PacSol} for $p_f$. Solution is shown as dashed magenta line over the raw waveform, and the inset shows a detail of the peak negative pressure portion of the waveform. (Panel 4) Scale $p_f$ to match known peak negative (threshold) pressure (magenta dashed line), and fit scaled $p_f$ with a continuous analytic pulse, $p_{ac}$ (Eq.\ \ref{eq:Pac}, black solid line). (Panel 5) Use $p_{ac}$ to obtain numerical radius vs.\ time solutions.}
 \label{fig:Pac}  
\end{figure}

Our proposed procedure for inferring the acoustic forcing waveform in histotripsy simulations is outlined in Fig.\ \ref{fig:Pac}. 
Our approach is premised on the notion that a nucleus does not become active until the empirically--determined intrinsic cavitation threshold (i.e., a peak rarefaction pressure of $-24$ MPa) is reached. It follows that the preceding portion of the waveform can be assumed to have no bearing on the nucleus. Furthermore, due to the rapid and large bubble growth, the portion of the pressure waveform following the large tension is no longer representative once a bubble has cavitated, thus suggesting the use of a half cycle. The setup begins with a representative bubble radius vs.\ time data set and the measured histotripsy forcing waveform shown in panel one. This raw waveform is deduced from voltage measurements made with a fiber--optic hydrophone at a pressure of $-18$ MPa (to avoid hydrophone damage) and exhibits electrical noise and nonlinear distortion. Under the constraint of producing single bubbles, the peak negative pressure in the focal region is equal to the experimentally--determined cavitation threshold of $-24$ MPa. The raw waveform shown in Panel 1 of Fig.\ \ref{fig:Pac} is scaled accordingly. Assuming that each bubble radius vs.\ time data set obeys a Rayleigh--Plesset--type equation, the data set is fitted with a smooth curve to obtain a hypothetical $R(t)$ solution (dashed black line in Panel 2 of Fig. \ref{fig:Pac}). A cubic splines method is used for the data fit to ensure continuity of the first and second derivatives of $R$, which are computed using central differences. One can then infer the acoustic forcing by solving for the forcing term in the Rayleigh--Plesset form of Eq.\ \ref{eq:RadEq}:
\begin{equation}
\begin{aligned}\label{eq:PacSol}
p_f(t) = p - \rho_{\infty}\left(R\ddot{R} + \frac{3}{2}\dot{R}^2\right)-p_{\infty}.
\end{aligned}
\end{equation}
\noindent
Use of the Rayleigh--Plesset equation avoids the time derivative of $p_f$ present in the Keller--Miksis formulations, and our results in Sect.\ \ref{sec:comp} confirm that ignoring compressibility to reconstruct the waveform is a valid treatment. The $p_f$ solution obtained using Eq.\ \ref{eq:PacSol} is shown in Panel 3 of Fig.\ \ref{fig:Pac} as the magenta dashed trace overlay on the raw waveform. As seen in the inset showing finer detail of the boxed region, the peak of the $p_f$ trace aligns with the peak negative pressure segment of the measured histotripsy pulse but is clearly of lower amplitude. The discrepancy between the amplitudes of the measured waveform and $p_f$ solution primarily due to the lack of experimental data (radius vs.\ time measurements) in the early stages of the nucleation process. In this modeling approach, the absence of radius vs.\ time data during early growth effectively starts from a $10$--micron nucleus. However, both experimental and theoretical evidence suggest that true threshold nuclei are on the order of nanometers \citep{maxwell2013probability,vlaisavljevich2015effects,vlaisavljevich2016effects}, and nucleus size is a key parametric uncertainty \citep{mancia2020measurements}. Instead, as in the case of the raw waveform, one can confidently take the peak negative pressure of the waveform to be equivalent to the acoustic cavitation threshold of $-24$ MPa.  We thus suggest scaling the $p_f$ solution such that its maximum tension is equal to this measured threshold. Panel 4 shows the magenta dashed trace corresponding to linear scaling of the $p_f$ solution. The scaled $p_f$ solution is then readily fit by a common approximation for histotripsy forcing, $p_{ac}$, shown in Panel 4 as the solid black trace and expressed analytically below:
\begin{equation}
\begin{aligned}
    p_{ac}(t)= 
\begin{cases}
    p_A\left(\frac{1+cos[\omega(t-\delta)]}{2}\right)^{\eta},&
      |t-\delta|\leq \frac{\pi}{\omega},\\ 0, & |t-\delta|>
      \frac{\pi}{\omega}.
\end{cases}
\label{eq:Pac}
\end{aligned}
\end{equation} 

\noindent
The measured peak pressure corresponds to $p_A = -24$ MPa in this analytic expression while the frequency, $f$, of the experimental waveform appears as $\omega = 2\pi f$ ($f = 1$ MHz) in Eq.\ \ref{eq:Pac}. The parameter $\delta$ is an arbitrary time delay, typically taken to be $5$ $\mu$s. The dimensionless fitting parameter $\eta$ is typically chosen to match the curvature of the peak tensile portion of a measured histotripsy waveform. Previous studies have used this analytic expression with $\eta = 3.7$ \citep{mancia2017predicting,mancia2019modeling,vlaisavljevich2015effects,vlaisavljevich2015effectsb,vlaisavljevich2016effects,vlaisavljevich2016visualizing}, a value deduced from a series of experimental waveforms with frequencies ranging from $0.345$--$3.0$ MHz \citep{vlaisavljevich2015effects}. The rising portion of the analytic waveform with $\eta = 3.7$ fits the scaled $p_f$ with an accuracy $>99.9 \%$. Thus, our $p_f$ solution inferred independently from experimental radius vs.\ time data and scaled to threshold amplitude is consistent with a common analytic approximation of the histotripsy waveform, supporting the validity of this analytic waveform for histotripsy simulations. The raw waveform is noticeably broader than both the $p_f$ solution and the fitted $p_{ac}$ waveform. This discrepancy likely reflects nonlinear broadening of the waveform that reaches the hydrophone tip, which could be more significant than broadening of the driving pressure experienced by an individual nucleus in the focal region.  The experimental histotripsy waveform capable of producing a single bubble at threshold is highly reproducible with negligible shot--to--shot variability, but a relatively wide range of $\eta$ produce analytic waveforms that fit the scaled $p_f$ solution obtained from radius vs.\ time data with an accuracy of at least $99.8\%$ (See Sect.\ \ref{sec:discuss}). Given relative insensitivity of numerical results to changes in $\eta$, nucleus size is considered the primary source of parametric uncertainty in simulations. Nucleus size is inferred by iterating over the initial radius used for simulations until a best fit simulation is found (Panel 5 of Fig.\ \ref{fig:Pac}). While this approach is applicable to any experimental data set, those with the greatest number of radius measurements spaced closely in time ensure a reliably smooth initial data fit.

\subsection{Compressibility Effects}
\label{sec:comp}

\begin{figure}
  \includegraphics[width=\linewidth]{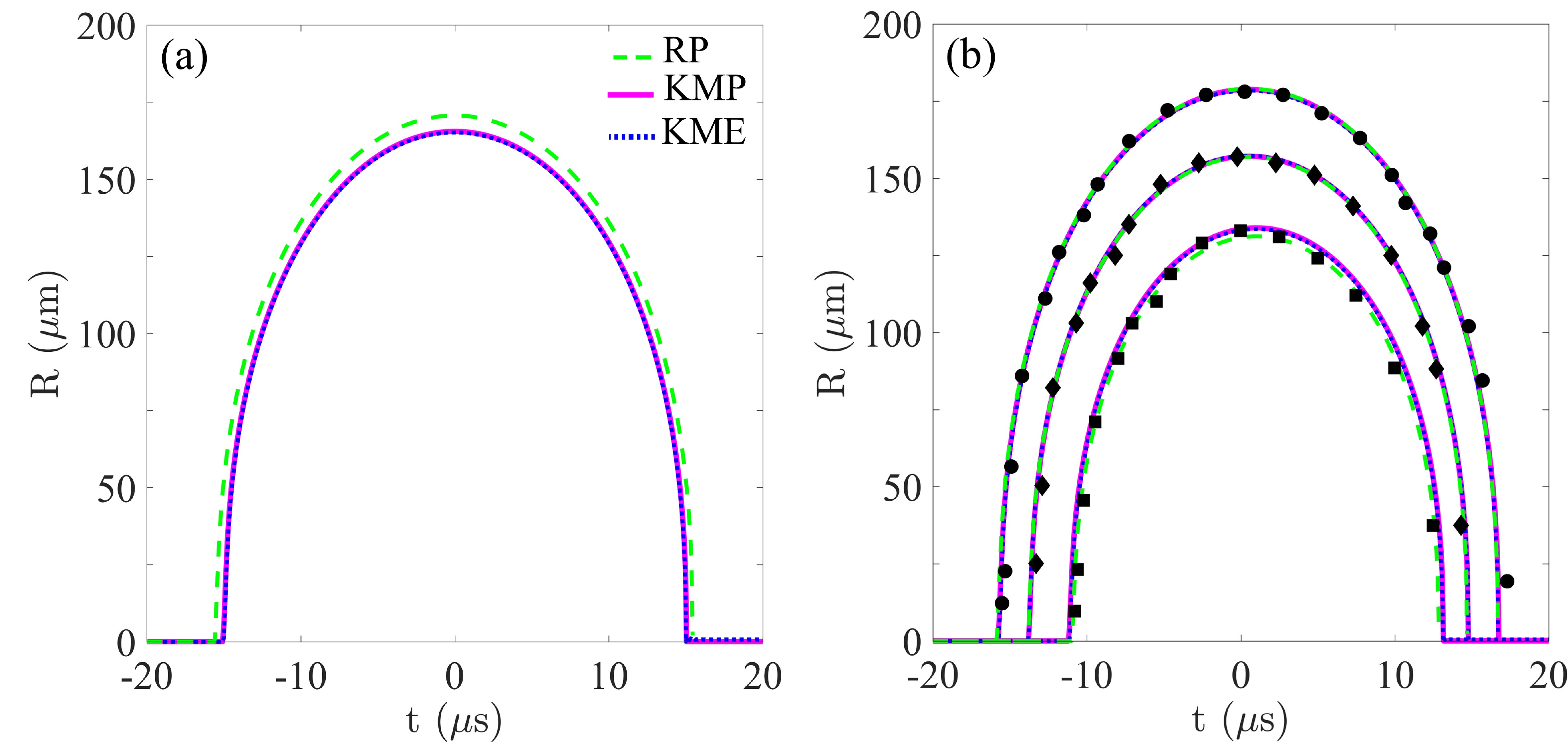}
  \centering
  \caption{(a) Radius vs.\ time simulations obtained with the cold liquid assumption for thermal effects and the Rayleigh--Plesset (RP), Keller--Miksis with pressure (KMP), and Keller--Miksis with enthalpy (KME) models for compressibility all initialized with a fixed $R_0$ of $3$ nm.  (b) The same models initialized with the optimized $R_0$ for each of three representative data sets: $R_{max}^{\mu - \sigma}$ (squares), $R_{max}^{\mu}$ (diamonds), and $R_{max}^{\mu + \sigma}$ (circles). The optimized $R_0$ values and NRMSE organized by data set and compressibility model are shown in Table \ref{table:RadEx}. }
  \label{fig:RadRvT}
\end{figure}

\begin{figure}
  \includegraphics[width=\linewidth]{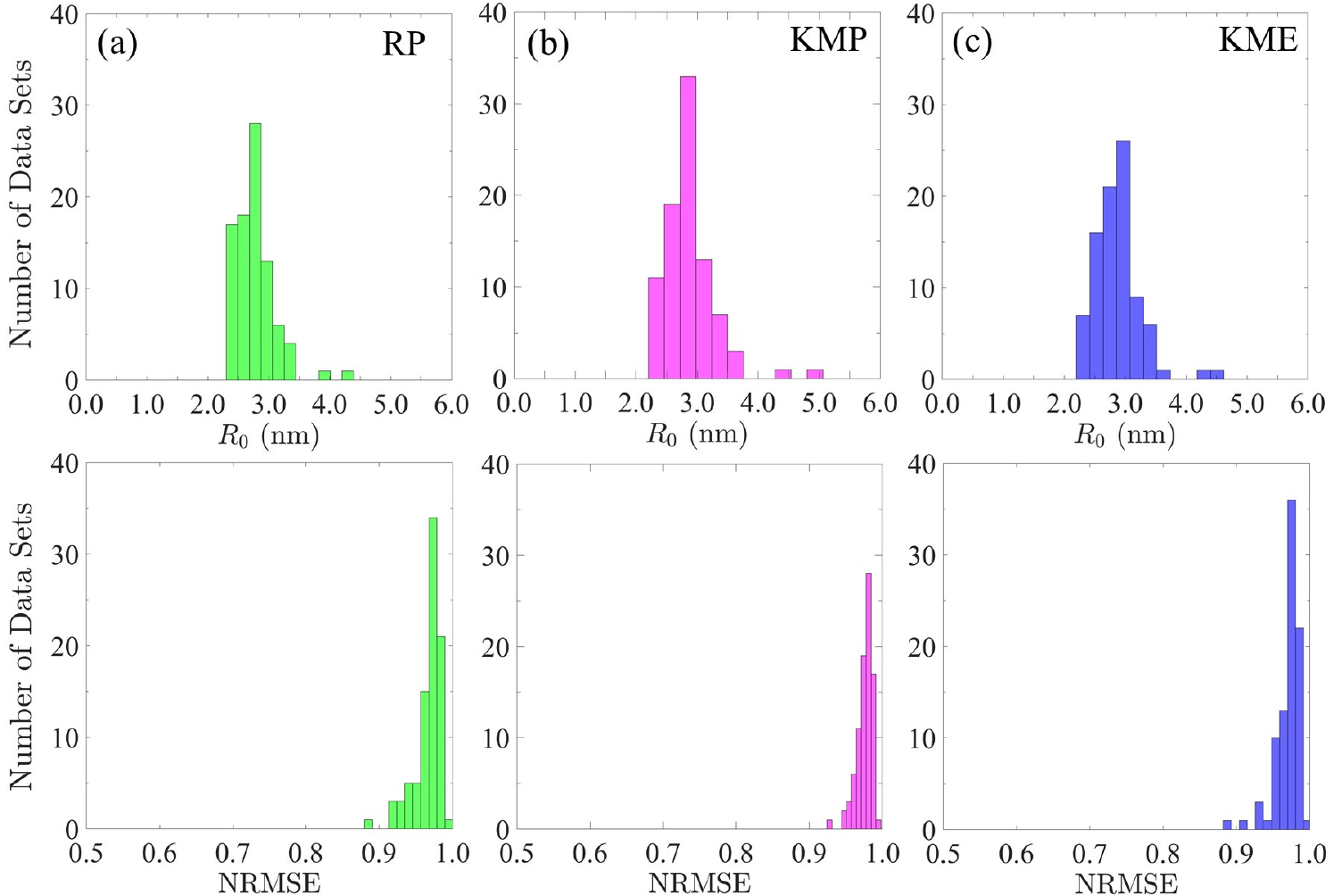}
  \centering
  \caption{Optimized $R_0$ distributions (top row) and NRMSE distributions (bottom row) obtained by applying a cold liquid assumption with each compressibility model to $88$ experimental data sets.}
  \label{fig:RadDist}
\end{figure}

Radial dynamics of a histotripsy bubble are modeled using the Rayleigh--Plesset equation (RP), the Keller--Miksis equation with pressure (KMP), and the Keller--Miksis equation with enthalpy (KME) forms of Eq.\ \ref{eq:RadEq}. The radius vs.\ time behavior produced by each radial dynamics model combined with the cold liquid assumption for a fixed $3.00$ nm initial radius is shown in Fig.\ \ref{fig:RadRvT}(a).  Consistent with expectations, the RP form results in a larger maximum radius due to the absence of compressibility as acoustic energy is radiated away. The KMP and KME forms both exhibit damping and differ from each other to a lesser degree, with the enthalpy form achieving a negligibly smaller maximum radius than the pressure form. Figure \ref{fig:RadRvT}(b) shows the optimal radius vs.\ time simulations obtained with each model by iterating $R_0$ for three representative data sets. The NRMSE and optimized $R_0$ obtained for each compressibility model applied to these data sets are shown in Table \ref{table:RadEx}.  The numerical radius vs.\ time results obtained with each model show good agreement with experimental data sets and considerable overlap with each other when initialized with the optimal $R_0$ for a given model and data set. The radial dynamics models differ most when applied to the $R_{max}^{\mu - \sigma}$ data set. The NRMSE ranges from $0.950$ for the Rayleigh--Plesset model applied to the $R_{max}^{\mu - \sigma}$ data set to $0.984$ for the Keller--Miksis models applied to the $R_{max}^{\mu + \sigma}$ data set.  The best agreement between experiment and simulation is achieved for the $R_{max}^{\mu + \sigma}$ data set regardless of the model used. Use of KMP or KME results in better agreement with experiments than using the RP form. The KMP model is slightly superior to KME for the $R_{max}^{\mu - \sigma}$ and $R_{max}^{\mu}$ data sets, though the KME model achieves agreement of $0.95$ or greater in more data sets than KMP. The optimized $R_0$ values are smallest for RP and larger for the two compressible models. Complete distributions of the NRMSE and optimized $R_0$ for $88$ data sets obtained using the cold liquid assumption with each of the compressibility models are shown in Fig.\ \ref{fig:RadDist}. The NRMSE and optimized $R_0$ distributions for each model and all data sets show considerable overlap. 

\begin{table*}[b]
\caption{NRMSE associated with the cold liquid assumption combined with each of three compressibility models. Models are applied to three representative data sets shown in Fig.\ \ref{fig:RadRvT}: $R_{max}^{\mu - \sigma}$ (squares), $R_{max}^{\mu}$ (diamonds), and $R_{max}^{\mu + \sigma}$ (circles). Mean optimized $R_0$ in nanometers is indicated in parentheses.}
\centering
\label{table:RadEx}
\begin{tabular}[width=\textwidth]{ c c c c } 
&\multicolumn{3}{c}{Compressibility Models}\\
\hline
Data Set & Rayleigh--Plesset & KM with Pressure & KM with Enthalpy\\
\hline 
$R_{max}^{\mu - \sigma}$ & 0.950 (2.39) & 0.969 (2.42) & 0.965 (2.45)  \\
$R_{max}^{\mu}$   & 0.973 (2.69) & 0.979 (2.78) & 0.974 (2.80) \\
$R_{max}^{\mu + \sigma}$   & 0.982 (3.30) & 0.984 (3.52) & 0.984 (3.52)  \\
  \hline\hline 
\end{tabular}
\end{table*}

\subsection{Thermal Effects}

\begin{figure}[t]
  \includegraphics[width=\linewidth]{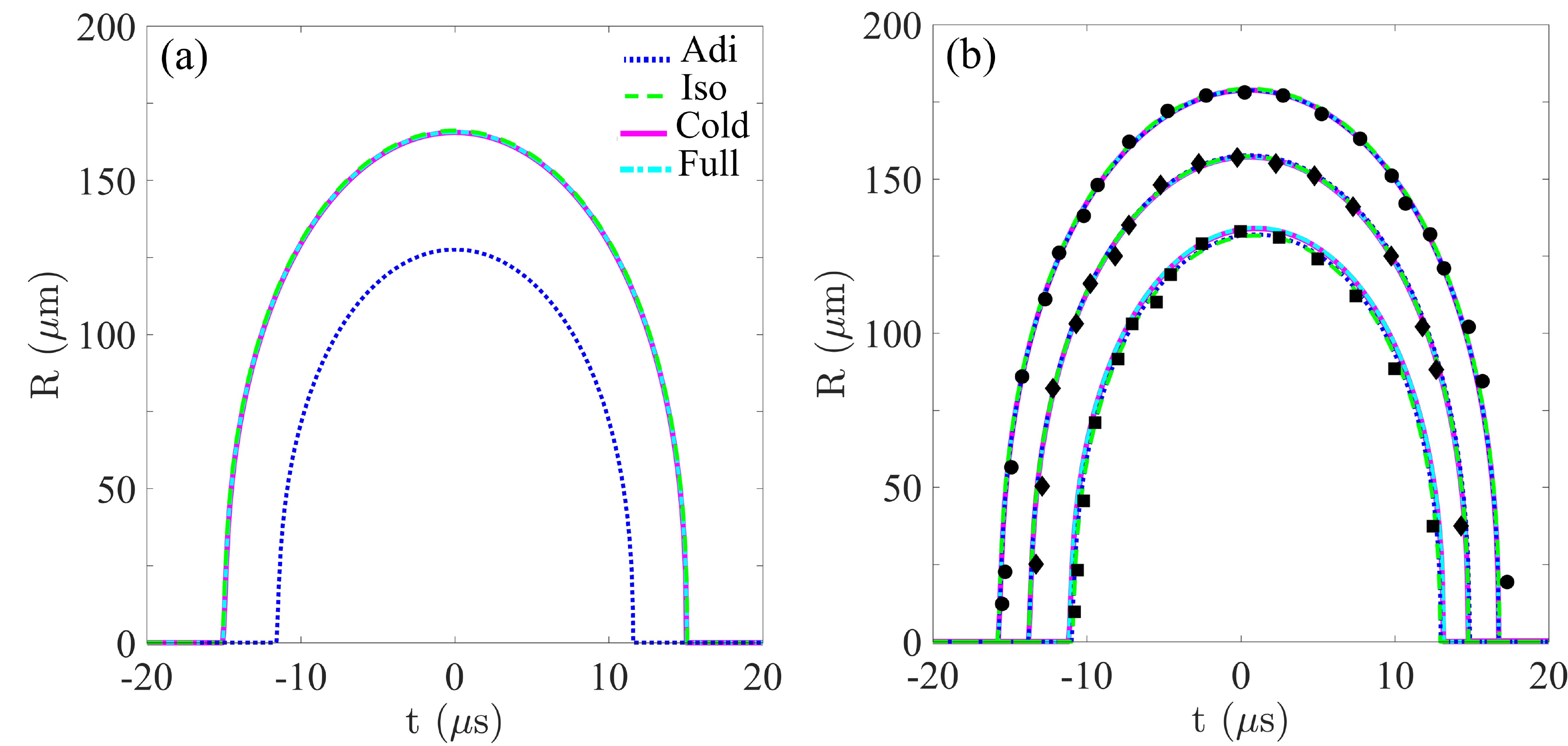}
  \centering
  \caption{(a) Radius vs.\ time simulations obtained with the KMP equation for compressibility effects and the adiabatic polytropic approximation (Adi), isothermal polytropic approximation (Iso), cold liquid assumption (Cold), and full thermal (Full) models all initialized with a fixed $R_0$ of $3.0$ nm.  (b) The same models initialized with the optimized $R_0$ for each of three representative data sets: $R_{max}^{\mu - \sigma}$ (squares), $R_{max}^{\mu}$ (diamonds), and $R_{max}^{\mu + \sigma}$ (circles). The optimized $R_0$ values and NRMSE organized by data set and thermal model are shown in Table \ref{table:ContEx}.}
  \label{fig:ContRvT}
\end{figure}

\begin{figure}
  \includegraphics[width=\linewidth]{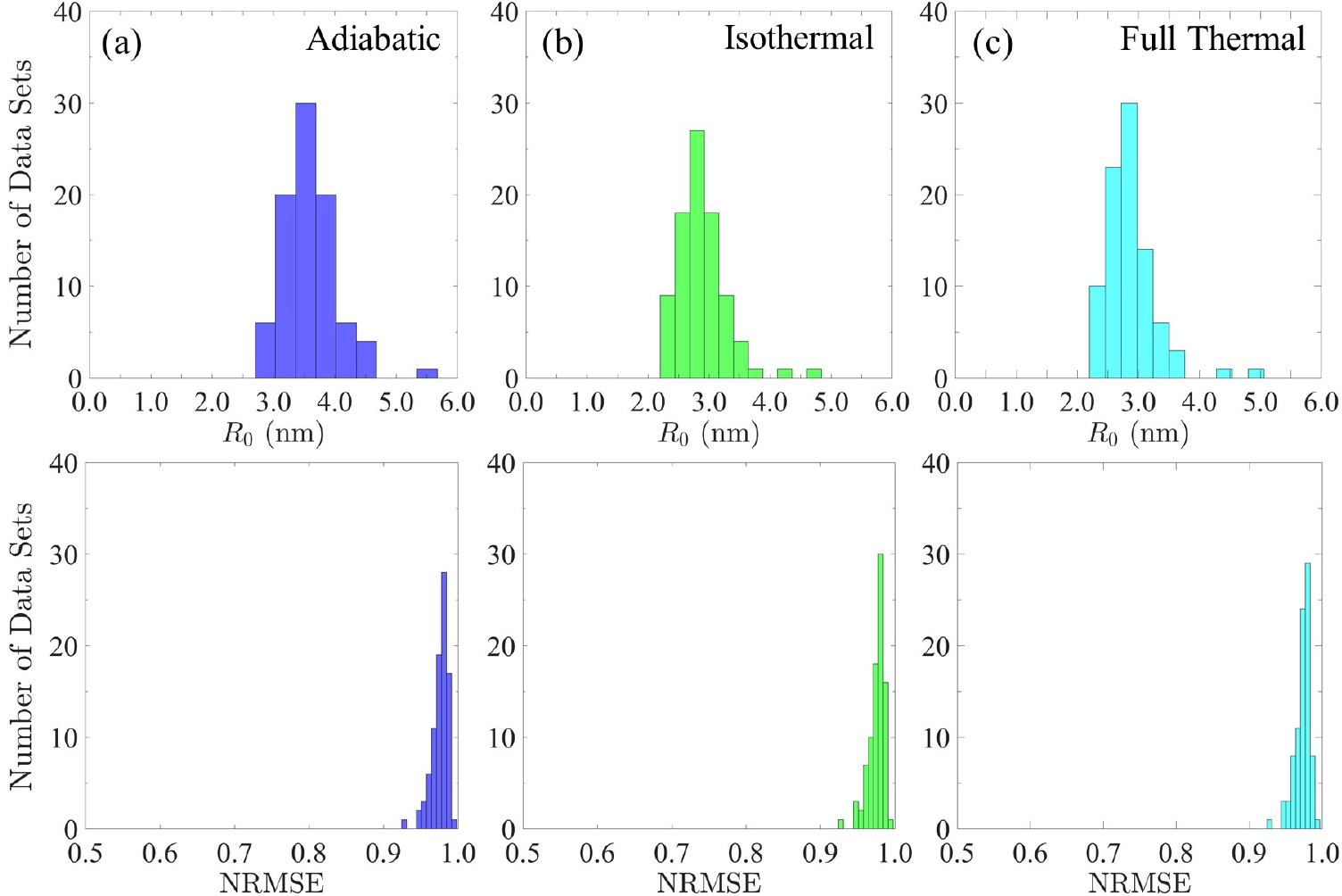}
  \centering
  \caption{Optimized $R_0$ distribution (top row), and NRMSE distribution (bottom row) obtained by applying the KMP equation with the (a) adiabatic polytropic approximation, (b) isothermal polytropic approximation, and (c) full thermal models for bubble contents to $88$ experimental data sets. The cold liquid assumption distributions are shown in Fig.\ref{fig:RadDist}(b).}
  \label{fig:ContDist}
\end{figure}

\begin{table*}[b]
\caption{NRMSE obtained using the KMP equation and each model for thermal effects. Models are applied to three representative data sets: $R_{max}^{\mu - \sigma}$ (squares), $R_{max}^{\mu}$ (diamonds), and $R_{max}^{\mu + \sigma}$ (circles) maximum radii. Mean optimized $R_0$ in nanometers is indicated in parentheses.}
\centering
\label{table:ContEx}
\begin{tabular}[width=\textwidth]{ c c c c c } 
&\multicolumn{4}{c}{Thermal Models}\\
\hline
Data Set & Adiabatic Polytropic & Isothermal Polytropic & Cold--Liquid & Full Thermal\\
\hline 
$R_{max}^{\mu - \sigma}$ & 0.969 (3.06) & 0.969 (2.42) & 0.969 (2.42)  & 0.961 (2.44)\\
$R_{max}^{\mu}$ & 0.979 (3.53) & 0.979 (2.78) & 0.979 (2.78) & 0.974 (2.78)\\
$R_{max}^{\mu + \sigma}$  & 0.984 (4.43) & 0.983 (3.51) & 0.984 (3.52) & 0.982 (3.52)  \\
  \hline\hline 
\end{tabular}
\end{table*}

The gaseous contents of a histotripsy bubble are modeled using the adiabatic polytropic approximation (Adi, Eq. \ref{eq:poly}, $\kappa = 1.4$), the isothermal polytropic approximation (Iso, Eq. \ref{eq:poly}, $\kappa = 1$), the cold--liquid assumption (Cold, Eqs. \ref{eq:pB} -- \ref{eq:Tm}), and the full thermal model (Full, Eqs. \ref{eq:pB} -- \ref{eq:Tm}). The radius vs.\ time behavior produced by each thermal model combined with the KMP equation for a fixed $3.00$ nm initial radius is shown in Fig.\ \ref{fig:ContRvT}(a).  As noted previously \citep{mancia2017predicting}, there is minimal distinction between the isothermal polytropic, cold--liquid assumption, and full thermal models for a histotripsy bubble. The most noticeable distinction is seen in the adiabatic polytropic case, which results in a significantly smaller maximum bubble radius for any given initial radius. Fig.\ \ref{fig:ContRvT}(b) shows optimized radius vs.\ time simulations for each thermal model coupled to the KMP equation for three representative data sets. The NRMSE and optimized $R_0$ obtained with each model applied to these data sets are shown in Table \ref{table:ContEx}.  The numerical results obtained with each model show good agreement with experimental data sets, and the radius vs.\ time traces overlap noticeably for all but the adiabatic polytropic approximation case. As in the compressibility models, differences in thermal models are most apparent when applied to the $R_{max}^{\mu - \sigma}$ data set. The NRMSE ranges from $0.961$ for the full thermal model applied to the $R_{max}^{\mu - \sigma}$ data set to $0.984$ for the adiabatic polytropic and cold--liquid models applied to the $R_{max}^{\mu + \sigma}$ data set.  As in the previous section, the best agreement between experiment and simulation is achieved for the $R_{max}^{\mu + \sigma}$ data set regardless of model. There are minimal differences in the NRMSE achieved with either the polytropic approximation or the cold--liquid assumption. The optimized $R_0$ values are nearly $1$ nm larger for the adiabatic polytropic approximation than for any other model; we note that the other models have near identical optimized $R_0$ values for the three data sets. The complete distributions of the NRMSE and optimized $R_0$ for $88$ data sets obtained using the KMP equation and each thermal model are shown in Fig.\ \ref{fig:ContDist}. Only the distributions for the polytropic approximation and full thermal models are shown in Fig.\ \ref{fig:ContDist} because the distributions with the cold--liquid assumption are already included in Fig.\ \ref{fig:RadDist}(b). The NRMSE distributions for each model are similar, with NRMSE $> 0.9$ for all cases. The optimized $R_0$ distributions are also similar for the isothermal polytropic, cold--liquid, and full thermal cases but with a noticeable shift to larger $R_0$ values for the adiabatic polytropic case.

\section{Discussion}
\label{sec:discuss}

\begin{table*}[t]
\caption{Mean NRMSE for each combination of compressibility and thermal model. Mean optimized $R_0$ in nanometers indicated in parentheses.}
\centering
\label{table:Summary}
\begin{tabular}[width=\textwidth]{ c c c c } 
&\multicolumn{3}{c}{Compressibility Models}\\
\hline
Thermal Models & Rayleigh--Plesset & KM with Pressure & KM with Enthalpy\\
\hline 
Adiabatic Polytropic & 0.967 (3.49) & 0.976 (3.63) & 0.974 (3.65)  \\
Isothermal Polytropic & 0.967 (2.77) & 0.975 (2.88) & 0.967 (2.89) \\
Cold--Liquid & 0.967 (2.77) & 0.976 (2.88) & 0.970 (2.89)  \\
Full Thermal & 0.967 (2.76) & 0.973 (2.88) & 0.967 (2.88) \\
  \hline\hline 
\end{tabular}
\smallskip
\end{table*}

Single--bubble models commonly used to simulate intrinsic--threshold histotripsy are found to be consistent with radius vs.\ time measurements obtained from bubbles nucleated by high--amplitude ultrasound. The combination of (i) single--bubble experimental data and (ii) a process for reconstructing the waveform enable us to validate a variety histotripsy single--bubble modeling approaches. Compressibility and thermal models of varying fidelities are applied to the single--bubble radius vs.\ time measurements in water, and the resulting mean NRMSEs and mean optimized initial radius sizes obtained from the $88$ experimental data sets with each combination of modeling approaches are summarized in Table \ref{table:Summary}. Notably, all modeling approaches achieve mean NRMSEs that are greater than $0.967$, and our results demonstrate a less than $1\%$ distinction between models for single--bubble compressibility and thermal effects applied with optimized $R_0$ values. Mean optimized initial radius sizes range from $2.76$ nm to $3.65$ nm, which are consistent with estimates for intrinsic nucleus sizes predicted previously \citep{mancia2020measurements,vlaisavljevich2015effects,maxwell2013probability}.  Given an appropriate choice of initial radius, all modeling approaches considered in this study appear to achieve similar predictions. Differences in mean $R_0$, while statistically significant, agree to within a tenth of a nanometer in all approaches except those with an adiabatic polytropic approximation. The significantly ($> 25\%$) larger optimal $R_0$ values obtained with the adiabatic polytropic approximation reflect this model's idealized neglect of heat transfer, which is important given the large heat capacity of water. Consistency among models employing the isothermal polytropic approximation, cold--liquid assumption, and full thermal model suggest that these are relatively interchangeable, again highlighting the fact that the temperature of the water does not change significantly during this rapid process.. These results also suggest that the isothermal polytropic approximation and cold--liquid assumptions are preferable to the adiabatic approximation given their consistency with the more complete physics included in the full thermal model. Given the accuracy of the isothermal model as well as the complexities and computational expense of the thermal models (full and cold liquid), it is reasonable to use isothermal to model histotripsy bubbles--as long as the initial conditions and forcing are appropriately determined. Greater guidance regarding best practices for modeling single--bubble dynamics in histotripsy requires considering the limitations of available data as well as the significance of model--based and parametric uncertainties during the bubble lifespan.

During early bubble growth, model validation is limited by an absence of experimental measurements at nucleation and the onset of rapid bubble expansion.  Although there is robust experimental evidence that histotripsy bubbles arise from preexisting, nanoscale nuclei \citep{mancia2020measurements,vlaisavljevich2015effects,maxwell2013probability}, the specific nucleus sizes are well below the spatial resolution of available measurement techniques. Given this expected scale of nucleus sizes, model limitations most pronounced during early bubble growth include the neglect of differences between local surface tension at the nanoscale bubble wall and its measured value in water as well as possible interactions between nuclei and ions or impurities in the water \citep{azouzi2013coherent}. The initial bubble radius or nucleus size, $R_0$, is a key parametric uncertainty in all of the models presented and must be inferred in an iterative fashion to initialize simulations. The resulting $R_0$ distributions are consistent with physical expectations that models accounting for thermal losses and acoustic damping will predict relatively larger $R_0$ values. For example, compressible models both predict significantly larger $R_0$ values than the incompressible RP equation. As previously noted, the adiabatic polytropic approximation, which assumes no heat transfer, predicts significantly larger $R_0$ values than the isothermal polytropic, cold--liquid, and full thermal models.  

Model validation in late collapse is similarly limited by an absence of experimental measurements. Perhaps the greatest challenge to the modeling approaches considered in this study and to lower--order models in general is observations of bubble breakup at collapse in histotripsy experiments \citep{duryea2015removal}. Bubble behavior at the instant of collapse to minimum radius is beyond the scope of the present study, but there are promising model considerations that could be justified when more data become available.  For example, bubbles often lose spherical symmetry near collapse \citep{ohl1998luminescence}, and consideration of more complex bubble geometries throughout the bubble lifespan \citep{prosperetti1977viscous,murakami2020shape} will ultimately be necessary for the prediction and modeling of bubble breakup.  Additionally, re--entrant jets can develop at bubble collapse near some heterogeneity or, in the setting of a histotripsy treatment, near another bubble in a bubble cloud \citep{callenaere2001cavitation}. Although future experimental work could potentially justify alternative treatments of bubble geometry, the incomplete understanding of bubble collapse physics in histotripsy remains a universal limitation of existing lower--order single--bubble models. 

The compressibility and thermal effects considered in this study are relevant throughout the bubble lifespan. In fact, the minimal ($< 1\%$) distinction seen among modeling approaches is likely explained by the absence of experimental data for the earliest stages of bubble growth and the latest stages of bubble collapse, when acoustic damping and thermal losses become most significant. Future experimental studies during these times could ultimately justify modifications to the models for compressibility and thermal effects presented here. For example, as mentioned previously, bubble wall velocity at the instant of collapse can exceed the range of validity for weakly compressible KMP and KME models and potentially may require higher fidelity in representing compressibility effects, such as the Gilmore model \citep{zilonova2018bubble}. Similarly, although the full thermal model provides a relatively complete treatment of heat transfer in bubble dynamics, even this approach neglects the dependence of thermal damping on driving frequency, nonuniform internal gas pressure, and the possibility of a nonconstant polytropic exponent \citep{prosperetti1977thermal,prosperetti1984bubble}. In addition, mass transfer effects are expected to be small because the mass boundary layer thickness remains small relative to the bubble radius throughout bubble expansion \citep{barajas2017effects}; however, these effects are also most significant during early bubble growth and late collapse \citep{bader2018influence,barajas2017effects} when experimental data is most lacking.

\begin{figure}
  \includegraphics[width=0.6\linewidth]{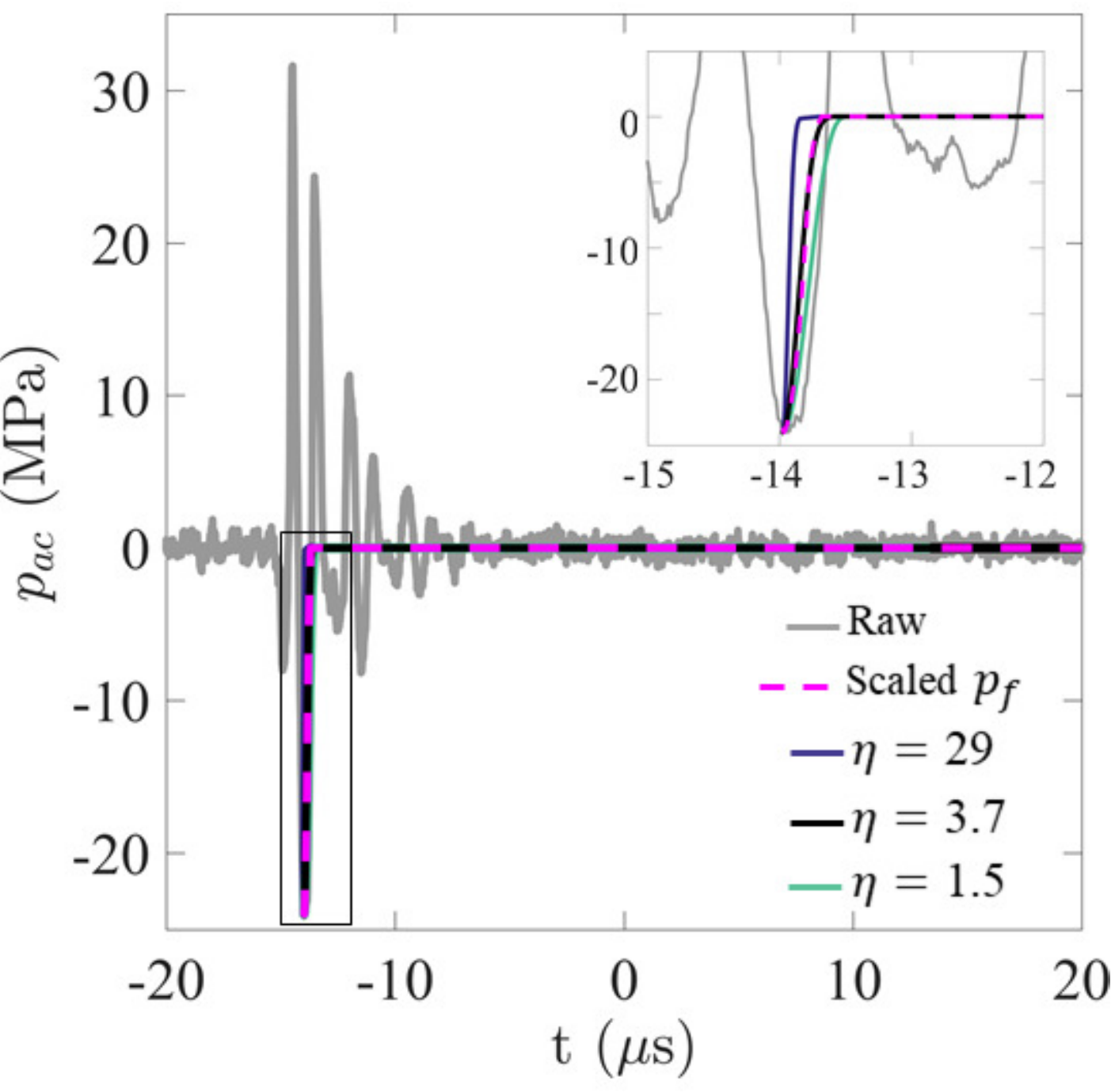}
  \centering
  \caption{Raw histotripsy waveform (gray) with overlay showing scaled waveform inferred from experimental data (dashed magenta) as well as the rising portion of analytic waveforms with fitting parameters $\eta = 3.7$ (black), $1.5$ (green), and $29$ (blue). Inset shows finer detail of boxed region.}
  \label{fig:WaveEx}
\end{figure}

The driving pressure remains a source of model--based uncertainty deserving of further attention. We infer the driving waveform from the experimental radius vs.\ time data by solving an inverse problem for $p_f$ (Eq.\ \ref{eq:PacSol}), scaling the resulting waveform to have an amplitude equal to the measured acoustic cavitation threshold, and fitting a symmetric analytic pulse, $p_{ac}$ (Eq.\ \ref{eq:Pac}), to this scaled $p_f$. The measured histotripsy waveform, the scaled $p_f$, and the rising portion of analytic pulses with select fitting parameters ($\eta=$ $3.7$, $1.5$, $29$) are shown in Figure \ref{fig:WaveEx}. There is minimal distinction in these waveforms at a timescale comparable to the lifespan of the bubble, but the measured waveform appears noticeably broader than the other waveforms in the inset. This apparent broadening of the measured waveform relative to scaled $p_f$ could be explained by nonlinear distortion of the 
acoustic waveform.  Specifically, nuclei in the focal region are expected to experience forcing that differs from the measured pulse reaching the hydrophone due to nonlinearities in the wave propagation. The inset also shows how the fitting parameter $\eta$ affects the shape of the analytic pulse (Eq.\ \ref{eq:Pac}). Excellent agreement between the scaled $p_f$ waveform and the rising portion of $p_{ac}$ with $\eta = 3.7$ is evident. In contrast, the rising portions of analytic pulses with $\eta = 1.5$ and $\eta = 29$ differ from the $p_{ac}$ solution to the same degree: $0.1$\%.  These results suggest that the analytic pulses with a wide range of $\eta$ values could be valid alternatives to the highly nonlinear measured waveform, but the precise relationship between these approximations and the measured waveform will require clarification with future experiments.  

Although the physical properties of water in Table \ref{table:Constants} are potential sources of parametric uncertainty throughout bubble growth and collapse, water is well--characterized under the room temperature conditions of the experiments, and minor variation in these parameters has a negligible effect on bubble dynamics \citep{mancia2019modeling}. Uncertainty in physical properties is a more significant consideration in tissue and other inhomogeneous materials, suggesting that simplified models (e.g.\ RP equation, polytropic approximations) could be useful in providing a limiting case solution when there is incomplete knowledge of these parameters. Likewise, all single--bubble models presented offer a resource--sparing, computationally--efficient alternative to higher--order methods. Despite some persistent model--based and parametric uncertainties most pronounced during early bubble growth and late bubble collapse, the consistency between single--bubble experiments and single--bubble modeling approaches supports the potential value of these models for high--throughput parameter studies. In particular, for an assumed nucleus size, these methods provide a rapid first--order approximation of maximum bubble radius, a key histotripsy damage metric \citep{mancia2017predicting,mancia2019modeling,bader2016predicting}. Moreover, application of our methods for characterizing the forcing waveform and inferring initial conditions from radius vs.\ time data can simplify efforts to validate models for bubble dynamics in viscoelastic materials. A similar approach could be used to determine appropriate constitutive models and viscoelastic properties for these materials under cavitation--relevant conditions \citep{estrada2018high}.

\section{Conclusions}

We demonstrated that single--bubble models commonly used to simulate intrinsic--threshold histotripsy are consistent with radius vs.\ time measurements obtained from bubbles nucleated in water by high--amplitude ultrasound. The combination of (i) single--bubble experimental data and (ii) a procedure for reconstructing the forcing experienced by a single bubble enabled us to validate a variety histotripsy single--bubble modeling approaches. Specifically, we compare the Rayleigh--Plesset equation and Keller--Miksis equations with pressure and enthalpy for compressibility effects as well as adiabatic and isothermal polytropic approximations, a cold--liquid assumption, and a full thermal model for thermal effects. Radius vs.\ time data obtained from single bubbles driven by a waveform with a single tensile phase is used to justify a common analytic approximation of the histotripsy waveform. We compare the optimized initial radius and associated NRMSE distributions obtained with each model applied to $88$ data sets and find that the single--bubble models considered in this study show excellent agreement with experimental results without significant distinction for the vast majority of histotripsy bubble growth and collapse in water. Optimal initial radius sizes for a given data set are comparable among models with the notable exception of significantly larger $R_0$ values obtained with the adiabatic polytropic model for thermal effects. This discrepancy suggests that heat transfer effects should not be neglected entirely and that the isothermal polytropic and cold--liquid models are acceptable alternatives to the full thermal model. Given an appropriately inferred initial radius, the minimal ($< 1\%$) distinction seen among modeling approaches is largely due to the absence of experimental data at the earliest stages of bubble growth and the latest stages of bubble collapse, when acoustic damping and thermal losses become most significant. However, the regime of model validity notably includes maximum bubble radius, a valuable histotripsy damage metric, suggesting this important quantity and other general features of histotripsy bubble dynamics can be adequately predicted with low--fidelity modeling approaches (e.g.\, Rayleigh--Plesset equation for an isothermal bubble) that require minimal computational resources. Future experimental measurements during early bubble growth and late bubble collapse could more definitively inform model choice. Similar methods could also be applied to validate models for histotripsy bubble dynamics in viscoelastic media. 

\appendix

%

\section*{Acknowledgments}
This work was supported by ONR Grant No.\ N00014-18-1-2625  (under  Dr.\ Timothy  Bentley).

\newcommand{\newblock}{}
\bibliographystyle{dcu}
\bibliography{HM_Bib}

%
%

\end{document}